\documentclass[a4paper,11pt,preprintnumbers]{article}

\usepackage{fullpage}
\usepackage[T1]{fontenc}

\usepackage{amssymb,amsmath}

\begin{document} 
	
\begin{titlepage}
	\thispagestyle{empty}
	\begin{flushright}
		
		\hfill{CPHT-RR037.032016 \\[1mm] DFPD-2016/TH/03} 
	\end{flushright}

	\vspace{35pt}
	
	\begin{center}
	    { \LARGE{\bf  On the origin of constrained superfields }}
		
		\vspace{50pt}
		
		{G.~Dall'Agata$^{1,2}$, E.~Dudas$^3$ and F.~Farakos$^{1,2}$}
		
		\vspace{25pt}
		
		{
		$^1${\it  Dipartimento di Fisica ``Galileo Galilei''\\
		Universit\`a di Padova, Via Marzolo 8, 35131 Padova, Italy}
		
		\vspace{15pt}
		
	    $^2${\it   INFN, Sezione di Padova \\
		Via Marzolo 8, 35131 Padova, Italy}
		
		\vspace{15pt}
		
		$^3${\it  Centre de Physique Th\'eorique, \'Ecole Polytechnique, CNRS, \\
		Universit\'e Paris-Saclay, F--91128 Palaiseau, France}
		}
		
		\vspace{40pt}
		
		{ABSTRACT} 
	\end{center}
	
In this work we analyze constrained superfields in supersymmetry and supergravity. 
We propose a constraint that, in combination with the constrained goldstino multiplet, consistently removes any selected component from a generic superfield.
We also describe its origin, providing the operators whose equations of motion lead to the decoupling of such components.
We illustrate our proposal by means of various examples and show how known constraints can be reproduced by our method. 
	
\vspace{10pt}

\bigskip

\end{titlepage}

\baselineskip 6 mm

\tableofcontents

\section{Introduction} 
\label{sec:introduction}

At energy scales well below the scale of supersymmetry breaking $\sqrt{f}$ supersymmetry is nonlinearly realized \cite{Volkov:1973ix}. 
While initially geometric methods were used for writing down the corresponding lagrangians \cite{Volkov:1973ix,Ivanov:1978mx}, it became clear soon that one could still efficiently employ superspace techniques if constrained superfields were introduced \cite{Rocek:1978nb,Casalbuoni:1988xh,Komargodski:2009rz}. 
Such superfields impose the decoupling of heavy states below the scale of supersymmetry breaking and implement nonlinear supersymmetry of effective low-energy actions in a novel way with respect to the more familiar introduction of explicit soft-breaking terms in lagrangians for the minimal supersymmetric standard model (MSSM). 
Used in a few works to define manifestly supersymmetric generalizations of MSSM  \cite{Antoniadis:2010hs}, they became more recently widely used in cosmology  \cite{Antoniadis:2014oya}--\cite{Antoniadis:2015ala} (for a recent review, see \cite{Ferrara:2015cwa}). 
This also revived a comprehensive study of constrained superfields in supergravity theories \cite{Dall'Agata:2014oka}, \cite{Dudas:2015eha}, \cite{Lindstrom:1979kq}-\cite{Kallosh:2016hcm}, which are the natural framework for building inflationary scenarios.

Even if we are mainly interested in models with constrained superfields to build effective theories, it is desirable to understand their origin from an ultraviolet (UV) perspective. 
First of all, it would single out stable microscopic frameworks, versus unstable ones, like models containing ghosts.  
Secondly, it would help clarifying the needed dynamics for removing the massive particles from the low-energy spectrum. 
Finally, it would make clear the generality and the naturalness of such constraints \cite{Dudas:2011kt}. 
In this paper we address this issue in a field-theoretical framework. 
Although explicit string constructions based on tachyon-free non-BPS systems are known since some time \cite{bsb} and their non-linear supersymmetry has been studied in detail \cite{dm}, at present their connection with the constrained  
superfield formalism is still unknown (for recent progress on the string theory description of the nilpotent goldstino see \cite{Kallosh:2015nia}). 

Our approach to generate the constraints is similar to the one used when passing from the linear to the non-linear $\sigma$-model. 
The original UV lagrangians have standard (linearly realized), spontaneously broken supersymmetry. 
Specific terms in the lagrangian generate large mass splittings in certain multiplets after supersymmetry breaking. 
At energy well below these masses, the (super)field equations of motion are dominated by these terms, which turn equations of motion into superfield constraints. 
Each such operator eliminates one field component. 
Several such operators, with large coefficients can eliminate several field components. 
Often the resulting multiple constraints can be combined into an equivalent single superfield constraint. 
We give several examples for chiral, vector and linear multiplets, with various components decoupled, in rigid supersymmetry and supergravity. 
To simplify the discussion, we assume that there is always a goldstino (chiral) multiplet, satisfying the standard polynomial constraint $X^2 = 0$.

In the simplest examples considered here, the UV actions are standard two-derivative supersymmetry/supergravity ones. 
However, in some examples, the UV operators needed to generate the desired constraints are of higher-derivative type, especially when considering constraints on the auxiliary fields.
This means that one should treat with care these models, because the original UV theory may be sick.

So far, the many superfield constraints that have been proposed seem to have no specific organizing principle behind them, 
and one has to find the appropriate constraint by a {\it trial and error} procedure. 
In this work we fill this gap and provide a simple organizing principle:
when the supersymmetry breaking sector has a goldstino superfield satisfying 
\begin{eqnarray}
\nonumber
X^2 = 0,
\end{eqnarray}
any constraint on another superfield $Q$ originates from
\begin{eqnarray}
\nonumber
X \overline X \, Q = 0.
\end{eqnarray}
This constraint removes the lowest component of $Q$ and is enough to explain the origin and the properties of all the known constraints in the literature, opening at the same time the way for many new possibilities.


\section{General constrained superfields and their origin} 
\label{sec:general_constrained_superfields_and_their_origin}

When supersymmetry is broken and non-linearly realized, it is known that one may remove various component fields from the spectrum by imposing supersymmetric constraints. 
Even though various examples are known, a systematic method to generate new constraints, and to reproduce all known ones as specific cases, is still lacking. 
In this section we aim to fill exactly this gap: we present a single superspace constraint  which, when imposed on a superfield, will remove from the spectrum its lowest component field. 
Moreover, imposing various constraints on the same superfield will result in removing additional component fields and may be equivalent to different known constraints. 
The structure of this generic constraint depends crucially on the properties of the supersymmetry breaking sector, which we shall review briefly. 

The breaking  of a global symmetry implies the existence of a massless goldstone mode. 
For supersymmetry this is the goldstino. 
In the simplest setup, the goldstino belongs to a chiral multiplet and has a scalar superpartner (the sgoldstino), which, once supersymmetry is broken,  acquires a non-supersymmetric mass. 
Decoupling the sgoldstino by giving it an {\it infinite} mass leads to  a non-linear realization of supersymmetry. 
In particular, the non-linear realization is induced by imposing on the SUSY-breaking chiral superfield $X$ the constraint \cite{Rocek:1978nb,Casalbuoni:1988xh,Komargodski:2009rz,Kuzenko:2010ef}
\begin{eqnarray}
\label{X2}
X^2 = 0 . 
\end{eqnarray} 
Whenever supersymmetry is broken by means of a non-trivial $F$-term $F \neq 0$, this constraint is solved by
\begin{eqnarray}
X = \frac{G^2}{2F} + \sqrt 2 \theta G + \theta^2 F . 
\end{eqnarray}
Actually, consistency does not require the existence of a decoupled massive scalar and therefore \eqref{X2} may be used more generally.
For instance, there are examples where this constraint does not appear as the infrared limit of the supersymmetry breaking sector \cite{Dudas:2011kt}, in case of generic couplings of $X$ to other decoupled scalar fields.
However, under specific assumptions on the UV theory, the constraint \eqref{X2} will generically hold.
In particular it will hold for a very heavy sgoldstino with small mass mixing to other fields. 

In this work we will always assume that \eqref{X2} is imposed on $X$, without further assumptions about the supersymmetry breaking sector, which will be described by the lagrangian
\begin{eqnarray}
\label{Xlagr}
{\cal L}_X = \int d^4 \theta X \overline X + \left\{ f \int d^2 \theta X + c.c. \right\} . 
\end{eqnarray} 
We can then construct generic effective theories when more heavy states have been integrated out by imposing constraints that remove the lowest component from a superfield $Q_L$, where the subscript ``L'' stands for a generic index labeling the Lorentz representation of the superfield\footnote{If the representation of the field $Q_L$ under the Lorentz group derives from the action of derivatives, one should be careful and check also the integrability of the solution to the constraint.}. 
This can be done by imposing 
\begin{eqnarray}
\label{XXQ}
X \overline X \, Q_\text{L} = 0 .
\end{eqnarray}
To remove more components from the same or from different supermultiplets, one has to impose several such constraints. 
As we will see, in most cases the constraint (\ref{XXQ}) can be understood as the decoupling of the specific component field, following from the introduction of a very large non-supersymmetric mass.  

A method to solve \eqref{XXQ} is to act on it with  various combinations of the superspace derivatives $D_\alpha$ and $\overline D_{\dot \alpha}$ and project the result to $\theta=\overline \theta=0$. 
This leads to various relations between the component fields of the superfield $Q_\text{L}$ and the component fields of the supersymmetry breaking sector. 
In fact the other conditions serve as consistency checks to the complete solution which is found by acting on \eqref{XXQ} with the maximum number of superspace derivatives, namely by taking 
\begin{eqnarray}
\label{maxD}
D^2 \overline D^2 \left( X \overline X Q_\text{L} \right) |= 0 ,
\end{eqnarray}
where the bar denotes the projection to $\theta = \overline \theta = 0$.
In other words, when imposing the constraint \eqref{XXQ} one simply has to solve \eqref{maxD} by expressing the component field $Q_\text{L}|$ in terms of the other component fields in the theory: 
\begin{eqnarray}
\label{maxDpr} 
Q_\text{L} = -2 \frac{\overline{D}_{\dot \beta} \overline{X} \, \overline{D}^{\dot \beta} Q_\text{L}}{\overline{D}^2 \overline{X}} 
- \frac{\overline{X} \, \overline{D}^2 Q_\text{L}}{\overline{D}^2 \overline{X}} 
-2 \frac{D^\alpha X D_\alpha \overline{D}^2 \left( \overline X Q_\text{L} \right) }{D^2 X \overline{D}^2 \overline{X}} 
- X \frac{D^2 \overline{D}^2 \left( \overline X Q_\text{L} \right) }{D^2 X \overline{D}^2 \overline{X}} . 
\end{eqnarray}
Once projected at $\theta = \overline \theta =0$, we get an expression for the component $Q_\text{L}|$ and all other conditions, which arise from (\ref{XXQ}) by acting with different combinations of $D_{\alpha}$ and $\overline{D}_{\dot \alpha}$ are identically satisfied.

In principle the component fields which reside in an unconstrained superfield give a reducible representation of the supersymmetry algebra, therefore we impose supersymmetric conditions to reduce the component content to an irreducible representation (for example  a chiral superfield has  $\overline D_{\dot \alpha} \Phi = 0$). 
It is easy to see that the supersymmetric conditions used to define the various superfields are respected by the constraint \eqref{XXQ}. 
As we show in the appendix, if we had acted on \eqref{maxDpr} with $\overline D_{\dot \gamma}$ without projecting to components we would find the identity $\overline D_{\dot \gamma} Q_\text{L}  = \overline D_{\dot \gamma} Q_\text{L} $. 
Similarly, if we had acted with $D_\gamma$ we would find $D_\gamma Q_\text{L}  = D_\gamma Q_\text{L} $. 
Clearly this property of the constraint will hold also when we act with more $D_\alpha$ or $\overline D_{\dot \alpha}$ and guarantees that the supersymmetric conditions on the superfield are not altered by the constraint. 
In particular it implies that the component fields all satisfy the same conditions as they did before we imposed the constraint \eqref{XXQ}. 
This is different from what happens for constraints that remove more than one component at once.

Now let us discuss the UV origin of the constraint \eqref{XXQ}. 
Assuming that \eqref{X2} holds, the constraint \eqref{XXQ} can be understood as the decoupling of the component field $Q_\text{L}|$, by taking some formal limit. 
We can illustrate this by a scenario where the supersymmetry breaking sector is appropriately coupled to the superfield $Q_\text{L}$, 
such that the component $Q_\text{L}|$ gets a non-supersymmetric mass. 
The total Lagrangian will have either the form 
\begin{eqnarray}
\label{inter1}
{\cal L} = {\cal L}_{X,Q_\text{L}} 
- \frac{m^2_{Q_\text{L}}}{2 f^2} \left\{ \int d^4 \theta X \overline X Q_\text{L}^2 + c.c. \right\} ,
\end{eqnarray}
or the form 
\begin{eqnarray}
\label{inter2}
{\cal L} = {\cal L}_{X,Q_\text{L}} 
- \frac{m^2_{Q_\text{L}}}{f^2}  \int d^4 \theta X \overline X Q_\text{L} \overline Q_\text{L},
\end{eqnarray}
depending on the properties of the superfield $Q_\text{L}$, and  ${\cal L}_{X,Q_\text{L}}$ stands for some supersymmetric Lagrangian containing the constrained superfield $X$ and $Q_\text{L}$. 
In the limit 
\begin{eqnarray}
m_{Q_\text{L}} \to \infty  
\end{eqnarray}
${Q_\text{L}}|$ decouples because it gets an {\it infinite} mass. 
The superspace equations of motion for  the superfield $Q_\text{L}$ will have a finite part  but also  a part which diverges. 
Due to the structure of the terms in \eqref{inter1} (or \eqref{inter2}) which give rise to the non-supersymmetric mass, 
the divergent part of the equations of motion will always identically vanish once we impose \eqref{XXQ} 
\begin{eqnarray} 
X \overline X Q_\text{L} =0 
\end{eqnarray} 
and viceversa, 
by requiring the divergent part to vanish, 
we will deduce the constraint \eqref{XXQ}. 
We will illustrate this with various examples in the next section.  

Let us end this general part with a comparison with the original approach in obtaining constrained multiplets. 
It was originally believed \cite{Komargodski:2009rz} that the constraints are unique and independent on the UV details, in particular on the masses of the decoupled supertpartners and more generally on the coefficients of the UV operators needed in the decoupling procedure.  
This would have indeed been desirable, because it would have implied the UV independence of the resulting constraints, like in the case of the Volkov--Akulov field alone. 
However, it was later realized \cite{Dudas:2011kt} that in the more general case where, in addition to the sgoldstino, superpartners in other multiplets are decoupled, the resulting constraints were generally modified. 
For this reason, here we take the simpler limit of infinite mass/coefficient for some operators. 
This should be a valid procedure in string theory examples  \cite{bsb,dm,Kallosh:2015nia}, where superpartners are just absent from the field-theory spectrum. 
From a field theory point of view it does imply specific UV assumptions on the dynamics, see e.g.~\cite{Dudas:2011kt}, \cite{Dudas:2016eej}.   


\section{General constrained superfields in global supersymmetry} 
\label{sec:general_constrained_superfields_in_global_supersymmetry}

\subsection{Constrained chiral superfields} 
\label{sub:constrained_chiral_superfields}

In this section we apply our general technique to constrained chiral superfields.
We start by considering models where a single component is removed from the spectrum and then move to more complicated examples involving multiple components.
We start by noting that the standard chiral constraint removing the scalar component from the chiral superfield $Y$, namely $XY = 0$ \cite{Brignole:1997pe}, is equivalent to the condition (\ref{XXQ}) introduced earlier:
\begin{equation}
	X \overline X \, Y = 0. \label{c1} 
\end{equation}
This is easily understood by recalling that the chiral superfield $\overline D^2 \overline X$  is nowhere vanishing because in the supersymmetry breaking sector we have 
\begin{eqnarray}
F  = - \frac14 \overline D^2 \overline X| = -f + \cdots \,.
\end{eqnarray} 
This means that the action of $\overline D^2$ on (\ref{c1}) gives a constraint equivalent to $XY = 0$ upon multiplication with $(\overline D^2 \overline X)^{-1}$.
Following the general procedure detailed in the previous section, the constraint (\ref{c1}) follows from a lagrangian containing the constrained goldstino multiplet, the $Y$ multiplet, and the coupling
\begin{equation}
-\frac{m_y^2  }{ f^2} \int d^4 \theta\, |X|^2 |Y|^2,
\end{equation}
which gives a non-supersymmetric mass to the lowest scalar component of $Y$. 
In the limit $m_y \to \infty$, the scalar is completely removed from the spectrum, leading to a constrained superfield. 
Indeed, in this limit, the superspace equations of motion for $Y$ have a $m_y$ dependent part which will diverge. 
Requiring the $m_y$ part to be identically vanishing yields  
\begin{equation}\label{eomYYY}
	\overline D^2 (|X|^2 \, \overline Y) =0,
\end{equation}
which is equivalent to \eqref{c1}.\footnote{Eq. \eqref{eomYYY} can be written as ${\overline{X}} D^2 (XY)=0$. After multiplication by $X$ and using the nilpotency of $X$ (which also implies $X D_\alpha X=0$), it becomes equivalent to \eqref{c1} upon multiplication with $(D^2 X)^{-1}$.}  

Another simple example is the decoupling of the fermion component field of $Y$. 
The term which generates the non-supersymmetric mass for $\sqrt2\, \chi_\alpha \equiv D_{\alpha} Y |$ has the form 
\begin{equation} \label{gero}
	- \frac{m_\chi}{2 f^2} \int \! d^4 \theta \Big{[}  |X|^2  D^{\alpha} Y D_\alpha Y + c.c. \Big{]}  
\end{equation}
and the divergent part of the superspace equations of motion in the limit $m_\chi \to \infty$ is 
\begin{eqnarray} \label{eomXDY}
\overline D^2 \left\{ D^\alpha ( |X|^2 D_\alpha Y ) \right\}  = 0 . 
\end{eqnarray} 
This is easily proved to be equivalent to
\begin{equation} \label{C}
	|X|^2 D_{\alpha} Y = 0 ,
\end{equation}
by acting on \eqref{eomXDY} with $\overline X D_\beta X$ and using the fact that $|D^2 X|^2 \neq 0$.
This constraint has been first proposed in \cite{Dall'Agata:2015lek} to consistently remove the fermion field in $Y$, while preserving a non-trivial $F$-term, which allows consistent general couplings to the rest of the matter multiplets\footnote{The operator (\ref{gero}) was proposed in Appendix C of 
\cite{Dudas:2011kt}. From our current discussion it is clear that in the infinite mass limit it leads to the constraint (\ref{C}) and not to the KS constraint (\ref{higgs}). This explains the puzzle mentioned there; the auxiliary field is indeed not removed. The existence of two different constraints removing the fermion from a chiral multiplet adds evidence to the non-uniqueness of the constrained superfields, even in the infinite mass limit.}.
This differs from the constraint
\begin{equation}\label{higgs}
	D_{\alpha} ( \overline X \, {\cal H}) = 0,
\end{equation}
which was proposed in \cite{Komargodski:2009rz} and removes both the fermion and the auxiliary component fields of the chiral superfield ${\cal H}$. 
Actually, we can prove that it is equivalent to imposing simultaneously two constraints of the form \eqref{XXQ}. 
First, if we multiply \eqref{higgs} by $X$ we recover 
\begin{eqnarray}
\label{firsthiggs2}
|X|^2 D_{\alpha} {\cal H} = 0 , 
\end{eqnarray}
which removes the fermion in ${\cal H}$.
Then, if we multiply \eqref{higgs} with $X D^\alpha$ we obtain
\begin{eqnarray}
\label{secondhiggs2}
  |X|^2 D^2 {\cal H} = 0,
\end{eqnarray} 
which removes the auxiliary field.
We can actually do more and prove that these two are also equivalent to (\ref{higgs}).
In fact, by acting on \eqref{firsthiggs2} with $D^2$ we get 
\begin{equation}
	(D^2 X) \overline{X} D_{\alpha} {\cal H} - \overline X D_{\alpha} X  D^2 {\cal H} = 0,
\end{equation}
and the second term is vanishing because of \eqref{secondhiggs2}, leaving us with the first term, which is equivalent to \eqref{higgs}. 
We now have a simple way to obtain (\ref{higgs}) from a lagrangian where only the goldstino multiplet is constrained.
We need to introduce large interaction terms of the form
\begin{equation} \label{hterms}
	 - \frac{m_h}{2 f^2} \int \! d^4 \theta \Big{[}  |X|^2  D^{\alpha} {\cal H} D_\alpha {\cal H} + c.c. \Big{]} 
	  - \frac{g_{F^H}}{f^2} \int \! d^4 \theta \Big{[} |X|^2 D^2 {\cal H} \overline D^2 \overline {\cal H} \Big{]}.
\end{equation} 
Note that if the chiral superfield $X$ was not nilpotent, the last term in \eqref{hterms} proportional to $g_{F^H}$ could  introduce ghosts into the theory. 
This is a non-trivial property of the decoupling procedure, and signal that such constraints could (but not necessarily) come from a sick UV theory.
We are interested once more in the dominant part of the ${\cal H}$ superspace equations of motion, in the limit of large $m_H/f^2$ and $g_{F^H}/f^2$ couplings. 
This is
\begin{equation}\label{Heom}
	\overline D^2 \left\{ \frac{m_h}{f^2} D^\alpha ( |X|^2 D_\alpha {\cal H} ) 
	- \frac{g_{F^H}}{f^2} D^2 (|X|^2 \overline D^2 \overline {\cal H}) \right\}  = 0 ,
\end{equation}
which, multiplied by $X \overline X$, produces
\begin{eqnarray}
|X|^2 |D^2 X|^2 \overline D^2 \overline {\cal H} = 0,
\end{eqnarray}
which is equivalent to \eqref{secondhiggs2}, and acted upon by $D_\beta X \overline X$ and using (\ref{secondhiggs2}) gives
\begin{eqnarray}
|X|^2 |D^2 X|^2  D_\beta {\cal H} = 0 ,
\end{eqnarray}
which implies \eqref{firsthiggs2}. 

It is interesting here to pause for a second and discuss an alternative way to impose a constraint on the $F$-term.
It is in fact clear that, in our setup, $|X|^2 D^2 {\cal H} = 0$ is equivalent to the antichiral constraint
\begin{equation}
	\overline X\, D^2 {{\cal H}} = 0.
\end{equation}
A suitable term that produces this constraint in the large mass limit is 
\begin{equation}\label{halternative}
	 - \frac{m_h}{2 f^2} \int \! d^4 \theta\, \overline X D^{\alpha} {\cal H} D_\alpha {\cal H} + c.c..
\end{equation}
The relevant part of the ${\cal H}$ equations of motion gives
\begin{equation}
	\overline D^2 D^{\alpha} (\overline X D_\alpha {\cal H}) = 0,
\end{equation}
which is equivalent to
\begin{equation}
	\overline D^2 (\overline X D^2 {\cal H}) = 0
\end{equation}
and, by multiplying with $\overline X$, to
\begin{equation}
	\overline X\, D^2 {{\cal H}} = 0.
\end{equation}
We will see in the next section that in the locally supersymmetric case one has to be careful in implementing the two options and in fact in some cases we are forced to choose the first over the second.

As a final example we study the constraint on the chiral superfield ${\cal A}$ that removes the imaginary part of the scalar, the fermion and the auxiliary field \cite{Komargodski:2009rz}:
\begin{equation}\label{axion}
	X ({\cal A} - \overline{\cal A}) = 0.
\end{equation}
It is a straightforward exercise to prove that (\ref{axion}) is equivalent to the following set of constraints
\begin{eqnarray}\label{ax-scalar}
 |X|^2 ({\cal A} - \overline{\cal A} )&=& 0,  \\[2mm]
\label{ax-fermion}
 |X|^2 \overline D_{\dot \alpha} \overline{\cal A} &=& 0, \\[2mm]
 \label{ax-Fterm}
 |X|^2 \overline D^2 \overline{\cal A} &=& 0.
\end{eqnarray}
This means that we can generate the constraint \eqref{axion} by means of three terms in the lagrangian, which generate non-supersymmetric masses for the component fields we remove:
\begin{equation}
	 \int \! d^4 \theta \Big{[} \frac{m^2_b}{2f^2} |X|^2 ({\cal A} - \overline{\cal A} )^2 
	- \frac{g_{F^A}}{f^2}  |X|^2 D^2 {\cal A} \overline D^2 \overline {\cal A} \Big{]} 
	- \frac{m_\zeta}{2 f^2} \int \! d^4 \theta \Big{[}  |X|^2  D^{\alpha} {\cal A} D_\alpha {\cal A} + c.c. \Big{]} . 
\end{equation} 
In the limit $m_b \to \infty$, $g_{F^A} \to \infty$ and $m_\zeta \to \infty$, the superspace equations of motion for the chiral superfield ${\cal A}$ are dominated by 
\begin{equation}\label{Aeom}
	\overline D^2 \left\{ \frac{m^2_b}{f^2} |X|^2 ({\cal A} - \overline{\cal A} )
	+\frac{m_\zeta}{f^2} D^\alpha ( |X|^2 D_\alpha {\cal A} ) 
	- \frac{g_{F^A}}{f^2} D^2 (|X|^2 \overline D^2 \overline {\cal A}) \right\}  = 0 .  
\end{equation} 
This reproduces the constraint (\ref{axion}) by looking at different projections: multiplication with $X \overline X$ gives \eqref{ax-Fterm}; the action with $D_\beta X \overline X$ gives \eqref{ax-fermion}; finally, using \eqref{ax-fermion} and \eqref{ax-Fterm} in (\ref{Aeom}) and then multiplying with $\overline X$ gives \eqref{ax-scalar}. 


\subsection{Constrained vector multiplets} 
\label{sub:constrained_vector_multiplets}

It is well known that in a theory with a vector multiplet coupled to the nilpotent goldstino superfield, a simple way to remove the gaugino $\lambda_{\alpha} = - i W_{\alpha}|$ from the spectrum is to impose \cite{Komargodski:2009rz}
\begin{equation}
X W_{\alpha} = 0 .  \label{XW1}
\end{equation}  
Clearly, such constraint is equivalent to 
\begin{eqnarray}
X \overline X \, W_\alpha = 0 .
\end{eqnarray} 
This form of the constraint can be generated by following our general procedure, introducing a large term of the form 
\begin{equation}
- \frac{m_\lambda}{2 f} \left( \int d^2 \theta \, X \,  W^{\alpha} W_{\alpha}  + c.c. \right) .  \label{v2}
\end{equation}
The field equations deriving from such term give
\begin{equation}
D^{\alpha} (X \, W_{\alpha}) + \overline D_{\dot \alpha} (\overline X \, \overline W^{\dot \alpha}) = 0 \ ,\label{v3}
\end{equation}
which, projected with $\overline X X D^{\beta}$, give
\begin{equation}
X \overline X W_{\alpha} = 0.\label{v4}
\end{equation}
We therefore see that \eqref{XW1} corresponds to the decoupling of the gaugino, due to its large mass.  

We could also consider other models where more components are removed from the spectrum by constraints of the form (\ref{XXQ}).
For instance, in the theory for a massive vector superfield 
\begin{eqnarray}
\label{Vmassive}
{\cal L } = \frac{m^2}{4} \int d^4 \theta (V - \Phi - \overline \Phi)^2 + \frac{1}{4 g^2} \left( \int d^2 \theta \, W^2 + c.c.  \right),  
\end{eqnarray}
gauge invariance is maintained if $\Phi \rightarrow \Phi + S$ together with $V \to V + S + \overline S$. 
This is a supersymmetric version of the Stueckelberg mechanism. 
Of course we may gauge fix $\Phi=0$, but then $V$ should not be written in the Wess-Zumino gauge. 
The component field spectrum of the Lagrangian \eqref{Vmassive} comprises a massive vector, a massive real scalar, two massive fermions with Dirac mass, a complex scalar auxiliary field, and a real scalar auxiliary field.  
All propagating fields have the same mass. 
The superspace constraint which removes the massive vector from the spectrum is 
\begin{eqnarray}
\label{XXV}
X \overline X [D_\alpha,\overline D_{\dot \alpha}] (V - \Phi - \overline \Phi)  = 0 
\end{eqnarray} 
and the term which gives a non-supersymmetric mass to the massive vector is 
\begin{eqnarray}
\frac{m^2_v}{16 f^2} \int d^4 \theta X \overline X [D_\alpha,\overline D_{\dot \alpha}] (V - \Phi - \overline \Phi)  [D^\alpha,\overline D^{\dot \alpha}](V - \Phi - \overline \Phi) . 
\end{eqnarray}


\subsection{Constrained real linear superfields} 
\label{sub:constrained_real_linear_superfields}

We conclude this section by analyzing constrained real linear superfields.
In this case we will give more details on the constraints and on their solutions, because they have not appeared previously in the literature.
A real linear superfield is defined by a real multiplet $L=L^*$ satisfying
\begin{eqnarray}
D^2 L = 0 = \overline D^2 L . 
\end{eqnarray} 
Its component fields are 
\begin{eqnarray}
\begin{split}
 \phi &= L | ,
\\[2mm]
\sqrt 2 \chi_\alpha  & = D_\alpha L |,
\\[2mm]
\sigma^m_{\alpha \dot \alpha} H_{m}    & = - \frac{1}{2} [D_\alpha , \overline D_{\dot \alpha}  ] L |.
\end{split}
\end{eqnarray}
Note that $H_m$ satisfies the constraint $\partial^m H_m = 0 $, which means that it is effectively the field strength of a real two-form 
\begin{eqnarray}
H_m =  \epsilon_{mnkl} \partial^n B^{kl} . 
\end{eqnarray}
We now present various constraints of the form (\ref{XXQ}), which remove some components of $L$.

To remove the real scalar we impose 
\begin{eqnarray}
\label{XXL}
X \overline X L = 0 ,
\end{eqnarray}
which leads to the equation
\begin{eqnarray}
\label{constrL}
L = - \frac{2 D^\alpha X D_\alpha L }{D^2 X} 
-D^2 \left\{ \frac{2 X \overline D_{\dot \alpha} \overline X \overline D^{\dot \alpha} L }{D^2 X \overline D^2 \overline X}  \right\} . 
\end{eqnarray}
It easy to see that the constraint $D^2 L=0$ is still satisfied. 
The property $L=L^*$ is not manifest, but it still holds. 
The component expression of \eqref{constrL} gives 
\begin{eqnarray}
\begin{split}
\phi =& \frac{\chi G}{F} + \frac{\overline \chi \overline G}{\overline F} + \frac{G \sigma^m \overline G}{2 F \overline F} H_m 
+ i \frac{G \sigma^m \overline \chi}{ F \overline F} \partial_m \left( \frac{\overline G^2}{2 \overline F} \right) 
- 4  i \frac{\overline G \overline \chi}{F \overline F} G \sigma^m \partial_m \overline G 
\\
& + i \frac{G \sigma^m \overline G}{2 F \overline F} \partial_m \phi 
-  \frac{G^2}{2 F^2 \overline F^2}  \partial_n \overline G \overline \sigma^n \sigma^m \overline \chi \partial_m \left( \frac{\overline G^2}{2 \overline F} \right) 
\\
& 
+ i \left( \frac{G^2}{2 F} \right) \partial_m \left( \frac{\overline G^2}{2 \overline F} \right) \frac{i \partial^m \phi +H^m}{F \overline F} 
+ i \frac{ G^2}{4 F^2 \overline F^2} \overline G \overline \sigma^m  \sigma^n \partial_n \overline G \left( i \partial_m \phi + H_m \right) 
\\
& - \left( \frac{ G^2}{2 F} \right) \partial^2 \left( \frac{\overline G^2}{2 \overline F} \right)  \frac{\overline G \overline \chi}{F \overline F^2} 
+  \frac{G^2 \overline G \overline \chi}{2 F^2 \overline F^3} \partial_m \overline G \overline \sigma^m \sigma^n \partial_n \overline G 
-i \frac{ G^2}{2 F^2 \overline F} \overline G \overline \sigma^m \partial_m \chi ,
\end{split}
\end{eqnarray}
which can be solved recursively to find $\phi$ in terms of the other component fields of the linear multiplet and the supersymmetry breaking sector. 
The leading terms in the expansion are 
\begin{eqnarray}
\phi = \frac{\chi G}{F} + \frac{\overline \chi \overline G}{\overline F} + \frac{G \sigma^m \overline G}{2 F \overline F} H_m + \cdots \,.
\end{eqnarray}
A  dynamical origin of the constraint \eqref{XXL} follows from the equations of motion generated by
\begin{eqnarray}
- \frac{m^2_\phi}{2 f^2}  \int d^4 \theta X \overline X L^2 ,
\end{eqnarray}
which give a non-supersymmetric mass to $\phi$ once the auxiliary field of the 
supersymmetry breaking sector $F$ is integrated out.
When $m_{\phi}$ is large, the scalar decouples and the constraint \eqref{XXL} follows.
Indeed we find that, in the $m_{\phi} \to \infty$ limit, the divergent term of the $L$ superspace equations of motion gives
\begin{equation}
	\overline{D}^2 D_{\alpha} (|X|^2 L) = 0 ,
\end{equation}
which, after acting with $\overline X D_\beta X$, becomes \eqref{XXL}. 

To remove the fermion we can impose the constraint 
\begin{eqnarray}
\label{XXDL} 
X \overline X D_\alpha L = 0 . 
\end{eqnarray}
$D_\alpha L$ is anti-chiral $D_\beta (D_\alpha L) = 0 $ and therefore the constraint simplifies to 
\begin{eqnarray}
\overline X D_\alpha L = 0 . 
\end{eqnarray}
When we project to the highest component we find 
\begin{eqnarray}
\label{constrchi}
\chi_\alpha = \frac{\overline G^{\dot \alpha}}{2 \overline F } \sigma^m_{\alpha \dot \alpha} \left( i \partial_m \phi - H_m \right) 
+ i \frac{\overline G^2}{2 \overline F^2}  \sigma^m_{\alpha \dot \alpha} \partial_m \overline \chi^{\dot \alpha} ,
\end{eqnarray}
while the lower component projections of the constraint are just consistency conditions. 
We  solve \eqref{constrchi} iteratively  to find 
\begin{eqnarray}
\begin{split}
\chi_\alpha &= \left( i \partial_{\alpha \dot \alpha} \phi  -H_{\alpha \dot \alpha} \right) \frac{\overline G^{\dot \alpha}}{2 \overline F}  
+ i \frac{\overline G^2}{2 \overline F^2} \partial_{\alpha \dot \alpha} \left( \frac{G_\rho (H^{\rho \dot \alpha} + i \partial^{\rho \dot \alpha} \phi)}{2 F} \right) 
\\
& - \frac{\overline G^2}{2 \overline F^2} \partial_{\alpha \dot \alpha} 
\left[ \frac{G^2}{2F^2} \partial^{\rho \dot \alpha} 
\left\{    
\left( i \partial_{\rho \dot \gamma} \phi  -H_{\rho \dot \gamma} \right) \frac{\overline G^{\dot \gamma}}{2 \overline F}  
+ i \frac{\overline G^2}{2 \overline F^2} \partial_{\rho \dot \gamma} \left( \frac{G_\gamma (H^{\gamma \dot \gamma} + i \partial^{\gamma \dot \gamma} \phi)}{2 F} \right) 
\right\} 
\right] . 
\end{split} 
\end{eqnarray}
Here we have used the notation $v_{\alpha \dot \alpha} = \sigma^n_{\alpha \dot \alpha} v_n$ 
and $v^{\alpha \dot \alpha} = \overline \sigma^{n  \dot \alpha \alpha} v_n $. 
To introduce a large non-supersymmetric mass for the fermion we can introduce the term 
\begin{eqnarray}
- \frac{m_\chi}{4 f^2} \int d^4 \theta X \overline X \left( D^\alpha L D_\alpha L + \overline D_{\dot \alpha} L \overline D^{\dot \alpha} L \right)  . 
\end{eqnarray}
By taking the formal limit $m_\chi \rightarrow \infty$  the fermion decouples, and the equations of motion remain finite if we impose \eqref{XXDL}. 
Indeed, the equations of motion in the large $m_\chi$ limit gives
\begin{equation}\label{DLeom}
	 \overline{D}^2 D^\beta \left\{ D_\alpha (|X|^2 D^{\alpha}L) 
	 + \overline{D}^{\dot \alpha} (  |X|^2 \overline D_{\dot \alpha} L)  \right\} = 0 ,
\end{equation}
which  we multiply by $|X|^2$ to deduce \eqref{XXDL}. 

For the real two-form $B_{kl}$ the situation changes  because $B_{kl}$ is a gauge field, 
and therefore it is protected by the gauge symmetry to be massless. 
If we insist on removing the two-form from the spectrum we have to embed it first  in a massive real linear multiplet. 
We note that the embedding into the massive real linear is essential {\it not} for the mass term it will give to the two-form, 
but rather for the extra degrees of freedom the two-form will get via the Stueckelberg mechanism by absorbing a U(1) vector. 
The real linear multiplet can be written with the help of a chiral prepotential ($\overline D_{\dot \beta} \, \tau_\alpha = 0$)  as 
\begin{eqnarray}
L = D^\alpha \tau_\alpha + \overline D_{\dot \alpha} \overline \tau^{\dot \alpha},
\end{eqnarray}
so that the superfield $\tau_\alpha$ has a gauge invariance 
\begin{eqnarray} \label{dt}
\tau_\alpha \to \tau_\alpha + i \, W_\alpha,
\end{eqnarray}
where the chiral superfield $W_\alpha$ is a vector field-strength superfield and $D_{(\alpha} \tau_{\beta)}$ contains the 2-form $B_{mn}$.
The manifest gauge invariant Lagrangian for the massive real linear multiplet is 
\begin{eqnarray}
{\cal L} = - \int d^4 \theta L^2  
- \frac{m^2_L}{2} \left(  \int d^2 \theta (\tau^\alpha - i \tilde W^\alpha)(\tau_\alpha - i \tilde W_\alpha) + c.c. \right),
\end{eqnarray}
where we have introduced a chiral superfield  
\begin{eqnarray}
\tilde W_\alpha = -\frac14 \overline D^2 D_\alpha \tilde V
\end{eqnarray}
and the gauge superfield  transforms as $\tilde V \to \tilde V + V$. 
In what follows we choose the gauge $\tilde V = 0$ to keep the formulas simple. 
To remove the two-form from the spectrum we impose the constraint 
\begin{eqnarray} \label{mB}
X \overline X \left( D_\alpha \tau_\beta   +  D_\beta  \tau_\alpha  \right) = 0 . 
\end{eqnarray}
The term which will give the large mass to the two-form is 
\begin{eqnarray}
- \frac{m^2_{B}}{2 f^2} \int d^4 \theta \Big{[} X \overline X \left( D_\alpha \tau_\beta   +  D_\beta  \tau_\alpha  \right)^2 + c.c. \Big{]}  . 
\end{eqnarray}
In the limit $m_B \to \infty$ the superspace equations of motion for $\tau_\alpha$ have a divergent term which reads 
\begin{eqnarray}
\overline D^2 D^\alpha \Big{[}  X \overline X \left( D_\alpha \tau_\beta   +  D_\beta  \tau_\alpha  \right) \Big{]} = 0 
\end{eqnarray}
 and once we multiply with $\overline X D_\gamma X$ we deduce \eqref{mB} (always keeping in mind that $D^2 X$ is invertible). 



\section{General constrained superfields in supergravity} 
\label{sec:general_constrained_superfields_in_supergravity}

Constrained superfields in supergravity have been discussed by various authors  \cite{Rocek:1978nb,Farakos:2013ih,Antoniadis:2014oya,Dall'Agata:2014oka}--\cite{Kallosh:2016hcm}, mainly reproducing the ones introduced in global supersymmetry.
Once more, our approach allows to consistently couple constrained superfields to supergravity via the introduction of the same general constraint as in global supersymmetry
\begin{eqnarray}
\label{sugraXXQ} 
X \overline X \, Q_\text{L} = 0 ,
\end{eqnarray} 
provided we have a nilpotent goldstino superfield $X$ in our model.
This includes all known examples and paves the way for many new possibilities.

Also in this case, this constraint can be recovered in supergravity by adding large non-supersymmetric mass contributions for the component we wish to remove.
This can be obtained once more by adding terms of the form
\begin{eqnarray}
{\cal L} =   \frac{m_{Q^2_\text{L}}}{16 f^2} 
\left[ \int d^2 \Theta \, 2 {\cal E}  \, (\overline {\cal D}^2 - 8 {\cal R}) \, 
  |X|^2 \, Q_\text{L}^2  \right] + c.c. 
\end{eqnarray}
or 
\begin{eqnarray} \label{sugramed2}
{\cal L} =   \frac{m_{Q^2_\text{L}}}{8 f^2} 
\left[ \int d^2 \Theta \, 2 {\cal E}  \, (\overline {\cal D}^2 - 8 {\cal R}) \, 
  |X|^2 \, Q_\text{L} \overline{Q}_\text{L}  \right] + c.c. 
\end{eqnarray}
depending on the properties of $Q_\text{L}$. 
Actually, if $Q_\text{L}$ is a chiral superfield, one may equivalently add terms of the form $m^2_{Q_\text{L}} |X|^2 |Q_\text{L}|^2$ to the K\"ahler potential, 
which give a large mass to $Q_\text{L}|$. 
When we take the formal limit $m_{Q_\text{L}} \to \infty$ and require the divergent part of the superspace equations of motion to be vanishing, we see that the constraint they imply is identically satisfied once we impose \eqref{sugraXXQ}.  
The applications this constraint can have are numerous, and we shall not discuss them here, but we will discuss various interesting examples containing chiral superfields. 

\subsection{Constraints on chiral superfields: removing physical fields }

The first example we wish to discuss is the elimination of the lowest complex scalar component $y$ from the chiral superfield $Y$. 
It is known that the constraint $X Y = 0$ can be consistently solved also in supergravity \cite{Dall'Agata:2015zla,Dall'Agata:2015lek} and eliminates $y$.
As in global supersymmetry, by multiplying it with $\overline X$ we get 
\begin{eqnarray}
\label{sugraXXY}
X \overline X Y = 0.
\end{eqnarray}
Viceversa, assuming that \eqref{sugraXXY} holds, we can multiply it with $\overline X \, \overline{\cal D}^2 $, 
which gives 
\begin{eqnarray}
\label{sugraalmostXY}
X \overline X \, \overline{\cal D}^2 \overline X \, Y = 0 ,
\end{eqnarray}
which is also equivalent to the previous one whenever supersymmetry is broken by the $X$ superfield.

To illustrate  the  origin of this constraint, 
we start from a supergravity theory coupled to $X$ and $Y$ via 
a K\"ahler potential   
\begin{equation}\label{KahlXY}
K = X \overline X + Y \overline Y - \frac{m_y^2}{f^2} |Y|^2 |X|^2 
\end{equation}
and a superpotential containing a non-trivial $F$-term for $X$. 
The superspace equations of motion for the chiral multiplet $Y$ are 
\begin{eqnarray}
(\overline {\cal D}^2 - 8 {\cal R} ) \text{e}^{-K/3} K_Y = 0 . 
\end{eqnarray}
Now to decouple the scalar $y$, we take the formal limit $m_y \to \infty$, 
and require the terms which diverge to be vanishing. 
This yields the constraint 
\begin{eqnarray}
\label{sugramuY}
 (\overline {\cal D}^2 - 8 {\cal R} ) \left\{  \text{e}^{-Y \overline Y /3}  (|X|^2 \overline Y - \frac{|X|^2 Y \overline Y^2}{3}  ) \right\}  = 0  . 
\end{eqnarray} 
We multiply with $\overline X$ to find  
\begin{eqnarray}
\text{e}^{-Y \overline Y /3}  |X|^2 \overline Y \left(1 -\frac{ Y \overline Y}{3} \right) = 0 
\end{eqnarray}
which is equivalent to \eqref{sugraXXY}, because we can multiply with $\frac{\text{e}^{Y \overline Y /3}}{ \left(1 -\frac{ Y \overline Y}{3} \right)} $. 

It is clear that, using a combination of constraints, one may remove from the spectrum of the theory a real scalar and the fermion in a single chiral multiplet 
\begin{eqnarray}
{\cal B} = b + i c + \sqrt 2 \Theta^\alpha \chi_\alpha + \Theta^2 F^B  . 
\end{eqnarray} 
By imposing on ${\cal B}$ the following constraints  
\begin{eqnarray}
\label{XXBB}
\begin{split}
|X|^2 ({\cal B} - \overline{\cal B}) =& 0 \ , 
\\
|X|^2 \, {\cal D}_\alpha {\cal B} =& 0 \ ,  
\end{split}
\end{eqnarray}
one may remove the real scalar $c$ and the fermion component of $\chi_\alpha$. 
To derive  these constraints as the decoupling of $c$ and $\chi_\alpha$, we can consider a K\"ahler potential of the form
\begin{eqnarray} 
K = X \overline X Z({\cal B}, \overline{\cal B}) + U({\cal B}, \overline{\cal B}) 
- \frac{m_c^2}{2 f^2} |X|^2 |{\cal B}-\overline{\cal B}|^2  \ , 
\end{eqnarray} 
together with a mediation term that gives a non-supersymmetric mass to the fermion 
\begin{eqnarray}
{\cal L} = \frac{m_\chi}{16 f^2} \left[ \int d^2 \Theta \, 2 {\cal E}  \, (\overline {\cal D}^2 - 8 {\cal R}) 
  |X|^2 \, {\cal D}^\alpha  {\cal B}    \,  {\cal D}_\alpha  {\cal B} \right] + c.c. 
\end{eqnarray}
and a superpotential containing a non-trivial $F$-term for $X$. 
The study of the divergent parts of the superspace equations of motion in the limits 
\begin{eqnarray}
\label{lm}
m_c \to \infty \ , \ m_\chi \to \infty  \ , 
\end{eqnarray}
leads to the constraints \eqref{XXBB}. 

\subsection{Constraints on chiral superfields: removing auxiliary fields }

Until now we have seen that supersymmetry and supergravity would give similar results. 
This does not happen when we start to remove auxiliary fields from the spectrum. 
We now illustrate this by considering a chiral superfield $Y$ and removing its auxiliary field $F^y$. 
As we have seen in global supersymmetry, this can be achieved by the introduction of one of the two different mediation terms: 
the second term inside \eqref{hterms} or \eqref{halternative}.  
The first option in supergravity becomes
\begin{eqnarray}
{\cal L}_\text{aux2} =   \frac{g_{F^y}}{8 f^2} \int \! d^2 \Theta \, 2 {\cal E} \, (\overline {\cal D}^2 - 8 {\cal R})   
\Big{[} |X|^2 {\cal D}^2 Y \overline{\cal D}^2 \overline Y \Big{]}  + c.c. \,.
\end{eqnarray}
The divergent part of the superspace equations of motion in the limit $g_{F^y} \to \infty$ give 
\begin{eqnarray} \label{aux1sugra}
|X|^2 {\cal D}^2 Y = 0  \  , 
\end{eqnarray}
which removes the auxiliary field $F^y$ from the spectrum. 
Alternatively, we could use the mediation term 
\begin{eqnarray} \label{Laux2sugra}
{\cal L}_\text{aux2} =  \frac{g_{F^y}}{8 f} \left[ \int d^2 \Theta \, 2 {\cal E}  \, (\overline {\cal D}^2 - 8 {\cal R}) 
  X \, \overline{\cal D}_{\dot \alpha}  \overline Y    \,  \overline{\cal D}^{\dot \alpha}  \overline Y \right] + c.c.  
\end{eqnarray}
which produces the constraint 
\begin{eqnarray} \label{aux2sugra}
X \overline{\cal D}^2 \overline{Y} = 0.
\end{eqnarray}
Clearly \eqref{aux2sugra} will give \eqref{aux1sugra} once we multiply with $\overline X$ and therefore it removes the auxiliary field from the spectrum. 
However, in contrast to the global case, once we act with $\overline{\cal D}_{\dot \alpha}$ on \eqref{aux2sugra} we find a non-trivial result 
\begin{eqnarray}
X \, {\cal R} \, \overline{\cal D}_{\dot \alpha} \overline{Y} = 0 . 
\end{eqnarray}
When supersymmetry is broken with vanishing vacuum energy, the lowest component of ${\cal R}$ always has a non-vanishing value, which is proportional to the gravitino mass. 
Therefore, in this setup one may consistently employ ${\cal R}^{-1}$ to find 
\begin{eqnarray}
X \, \overline{\cal D}_{\dot \alpha} \overline{Y} = 0  \ , 
\end{eqnarray}
which removes also the fermion $\chi_\alpha$ from the spectrum. 
This is an example of a constraint being imposed indirectly via {\it gravity mediation}. 
Indeed, if we turn to the exact component form of \eqref{Laux2sugra}, 
we find (in the gauge $G_\alpha=0$) 
\begin{eqnarray} \label{Laux2sugracomp}
\text{e}^{-1} {\cal L}_\text{aux2} =  \frac{g_{F^y}}{ f} \left[ F \, (\overline{F}^y)^2 + M \, F \, \overline{\chi}^2    \right] + c.c.  
\end{eqnarray}
which shows that the fermion will get a mass when for the supergravity auxiliary field $M$ we have $\langle M \rangle \ne 0$. 


\section{Summary and prospects}

In this work we have studied theories where the supersymmetry breaking sector is described by the nilpotent superfield $X$. 
We have shown that all known constraints on additional matter and gauge superfields are manifestations of a single generic constraint. 
We have also shown that all known constraints plus some new ones we derive, can be understood microscopically as arising from the decoupling of heavy states in the infinite mass limit or, for the case of auxiliary fields, in the infinite coupling limit of appropriate operators. 
For rigid supersymmetry and in all examples, for each component field removed there is a corresponding operator in the UV.  
In the case of supergravity, one operator can decouple several field components simultaneously and we gave an example where a single operator decouples simultaneously a fermion and an auxiliary field.  
   
It would be interesting to understand under which conditions the formal limit of infinite mass and couplings is a good approximation of  UV dynamics. From this perspective, our examples are not really microscopic UV models leading to the superfield constraints in the IR, but a parametrization of needed operators with large coefficients. We believe however that they are a first step in unravelling the required UV dynamics, by identifying the necessary ingredients needed in finding truly microscopic models.   

\section*{Acknowledgements}

We thank S.~Ferrara, A.~Kehagias, A.~Sagnotti and F.~Zwirner for valuable discussions.
G.D.~and F.F.~would like to thank the CPhT of the \'Ecole Polytechnique and CNRS for hospitality and support during the completion of this work.
This work was also supported in part by the MIUR grant RBFR10QS5J (STaFI) and by the Padova University Project CPDA119349.

\appendix

\section{Self-consistency of the constraint}

In this appendix we prove that the constraint \eqref{XXQ} removes no other component from the superfield $Q_\text{L}$, other than the lowest one. 
This is proved by showing that by acting with $D_\alpha$ or $\overline D_{\dot \alpha}$ on the condition constraining the $\theta = \bar \theta = 0$ component of $Q$, we obtain identities, and therefore we do not impose any further constraint on the higher components. 
From \eqref{XXQ} we have 
\begin{eqnarray}
X \overline{D}^2 \left( \overline X Q_\text{L} \right) = 0  \ , 
\end{eqnarray}
which gives 
\begin{eqnarray}
\label{A}
\overline{D}^2 \left( \overline X Q_\text{L} \right)  = -2 \frac{D^\alpha X D_\alpha \overline{D}^2 \left( \overline X Q_\text{L} \right) }{D^2 X} 
- X \frac{D^2 \overline{D}^2 \left( \overline X Q_\text{L} \right) }{D^2 X} \ . 
\end{eqnarray}
Then from \eqref{A} we find 
\begin{eqnarray}
\label{B}
Q_\text{L} = -2 \frac{\overline{D}_{\dot \beta} \overline{X} \, \overline{D}^{\dot \beta} Q_\text{L}}{\overline{D}^2 \overline{X}} 
- \frac{\overline{X} \, \overline{D}^2 Q_\text{L}}{\overline{D}^2 \overline{X}} 
-2 \frac{D^\alpha X D_\alpha \overline{D}^2 \left( \overline X Q_\text{L} \right) }{D^2 X \overline{D}^2 \overline{X}} 
- X \frac{D^2 \overline{D}^2 \left( \overline X Q_\text{L} \right) }{D^2 X \overline{D}^2 \overline{X}} \ , 
\end{eqnarray}
which is the form of the solution we shall use here to prove the consistency. 
By using \eqref{B} whenever $Q_L$ appears without derivatives in (\ref{A}), one can show that it is satisfied with no further assumptions. 
Then, by using \eqref{A} one can show that 
\begin{eqnarray}
\label{C2}
\overline D_{\dot \alpha} \left( -2 \frac{D^\alpha X D_\alpha \overline{D}^2 \left( \overline X Q_\text{L} \right)}{D^2 X} 
- X \frac{D^2 \overline{D}^2 \left( \overline X Q_\text{L} \right)}{D^2 X} \right) =  0 \ ,
\end{eqnarray}
by repeatedly replacing $\overline{D}^2 \left( \overline X Q_\text{L} \right)$ with the right hand side of \eqref{A}. 
This is the same as repeatedly replacing \eqref{B} into the left hand side of \eqref{C2} until  all terms vanish. 
We find for \eqref{B} that  
\begin{eqnarray}
 \overline D_{\dot \alpha} \left(  -2 \frac{\overline{D}_{\dot \beta} \overline{X} \, \overline{D}^{\dot \beta} Q_\text{L}}{\overline{D}^2 \overline{X}} 
- \frac{\overline{X} \, \overline{D}^2 Q_\text{L}}{\overline{D}^2 \overline{X}} 
-2 \frac{D^\alpha X D_\alpha \overline{D}^2 \left( \overline X Q_\text{L} \right) }{D^2 X \overline{D}^2 \overline{X}} 
- X \frac{D^2 \overline{D}^2 \left( \overline X Q_\text{L} \right) }{D^2 X \overline{D}^2 \overline{X}}  \right) =  \overline D_{\dot \alpha} Q_\text{L} \ , 
\end{eqnarray} 
where one has to use \eqref{C2}.  
We also find that  
\begin{eqnarray}
D_\beta \left(  -2 \frac{\overline{D}_{\dot \beta} \overline{X} \, \overline{D}^{\dot \beta} Q_\text{L}}{\overline{D}^2 \overline{X}} 
- \frac{\overline{X} \, \overline{D}^2 Q_\text{L}}{\overline{D}^2 \overline{X}} 
-2 \frac{D^\alpha X D_\alpha \overline{D}^2 \left( \overline X Q_\text{L} \right) }{D^2 X \overline{D}^2 \overline{X}} 
- X \frac{D^2 \overline{D}^2 \left( \overline X Q_\text{L} \right) }{D^2 X \overline{D}^2 \overline{X}} \right) =  D_\beta Q_\text{L}  \ , 
\end{eqnarray}
where one needs only the repeated use of \eqref{B}.  
This proves that we are removing only the component field $Q_\text{L}|$ from the superfield $Q_\text{L}$.

\end{document}